# Ice Formation via Deposition Mode Nucleation on Bare and Alcohol-covered Graphite Surfaces

Xiangrui Kong, Patrik U. Andersson, Erik S. Thomson, and
Jan B. C. Pettersson*

*Department of Chemistry, Atmospheric Science, University of Gothenburg, SE-412 96 Gothenburg, Sweden*

\* To whom correspondence should be addressed. E-mail: janp@chem.gu.se. Phone: +46 31 786 90 72. Fax: +46 31 772 13 94.

**Abstract**

Deposition of water on aerosol particles contributes to ice cloud formation in the atmosphere with implications for the water cycle and climate on Earth. The heterogeneous ice nucleation process is influenced by physico-chemical properties of the substrate, but the mechanisms remain incompletely understood. Here, we report on ice formation on bare and alcohol-covered graphite at temperatures from 175 to 213 K, probed by elastic helium and light scattering. Water has a low wettability on bare and butanol-covered graphite resulting in the growth of rough ice surfaces. In contrast, pre-adsorbed methanol provides hydrophilic surface sites and results in the formation of smooth crystalline ice; an effect that is pronounced also for sub-monolayer methanol coverages. The alcohols primarily reside at the ice surface and at the ice-graphite interface with a minor fraction being incorporated into the growing ice structures. Methanol has no observable effect on gas/solid water vapor exchange whereas butanol acts as a transport barrier for water resulting in a reduction in ice evaporation rate at 185 K. Implications for the description of deposition mode freezing are discussed.

## 1. Introduction

Clouds have important effects on the water cycle on Earth and on the radiation budget of the atmosphere. The formation of liquid cloud droplets is relatively well described by existing theory, while the formation of ice particles remains poorly understood. This is a major concern since it introduces uncertainties in the description of clouds and aerosols and limits our ability to model Earth's climate. In the atmosphere water droplets tend to remain in a super-cooled state far below the melting point of water with temperatures of about 235 K required to homogeneously freeze pure water cloud droplets.[1,2] At higher temperatures ice may form by heterogeneous ice nucleation involving insoluble aerosol particles. Four heterogeneous nucleation mechanisms have been identified:[1-3] 1) *deposition freezing* occurs when ice forms directly by water adsorption on surfaces, 2) *condensation freezing* occurs when liquid water first condenses and subsequently freezes, 3) *immersion freezing* occurs when an aerosol particle within a droplet induces freezing after droplet cooling, and 4) *contact freezing* occurs when an aerosol particle induces freezing when making surface contact with a droplet. Condensation, immersion and contact mode freezing are most important between approximately 235 K and 273 K when supercooled water droplets are metastable over long times. The lower temperature bound can only be approximated because aerosol physical and chemical properties, dissolved components, and dynamic processes in the atmosphere can all act to modify the nucleation rate.[1-4] Here we concentrate on deposition mode ice nucleation, which primarily contributes to ice nucleation at temperatures below approximately 243 K,[5] and is more important than the other heterogeneous ice nucleation modes at temperatures below 235 K. Deposition freezing is thus of particular relevance for clouds in the upper troposphere. Formation of cirrus clouds is expected to be dominated by homogeneous freezing in ascending air parcels,[6-9] but heterogeneous freezing may substantially modify cloud properties depending on aerosol particle concentrations.[6-11]

Aerosol particles that act as ice nuclei (IN) in the upper troposphere include mineral dust, soot, and metallic particles.[11-13] Mineral dust particles appear to be more efficient IN than soot and organic particles, while oxidized organic compounds have been shown to be better than reduced compounds.[14,15] Aerosol particle size plays a role and large particles may activate at a lower supersaturation than small particles.[2,10,16-18] However, in general only a minor fraction of existing particles activate and it has been difficult to provide causal links between ice nucleation and IN properties and composition. We focus on carbon-based systems, including carbon surfaces and surfaces modified by adsorption of organic compounds. Freshly emitted soot particles are generally hydrophobic and not efficient IN, but chemical and microphysical aging and adsorption of surface coatings may influence their ice-forming abilities.[19] For soot a wide range of ice nucleation characteristics have been observed in laboratory studies and depend on formation process, particle size and adsorbates.[20-25] Between 213 K and 233 K "lamp black" soot particles act as deposition nuclei only at high ice supersaturations near water saturation.[20] In contrast, soot particles with large specific surface areas and organic carbon mass contents of ~ 10%, generated from spark discharge, activate at low ice saturation ratios between 1.1 and 1.4.[21] In experiments with propane flame soot particles, ice nucleation was found to be most efficient for particles with an organic carbon mass content of ~ 5%, while high organic carbon contents of 30% and 70% lead to a marked suppression of nucleation efficiency.[24] The ice nucleation efficiency of soot particles from different combustion sources representing a range of physico-chemical properties has also been investigated, with hydrophobic soot types shown to be poor IN while nucleation was favored on oxidized hydrophilic soot of intermediate polarity.[23] The notable differences between the studies may be attributed to different surface characteristics including chemical composition, porosity, and specific surface area. These studies suggest that soot may be a significant source of atmospheric IN, although a lack of fundamental understanding makes extrapolations to atmospheric conditions speculative. Water binds weakly to the graphite plane[26,27] and water adsorption is therefore assumed to depend on the presence of hydrophilic adsorption sites.[28] This appears to describe soot particles that are cleaned by outgassing and heating, while ice nucleation on more realistic combustion particles is likely to be affected by the presence of condensable organic and inorganic materials.

A large number of condensable organic compounds in the atmosphere may potentially affect ice nucleation. Alcohols are one group of organic compounds that are well known to have a significant impact on ice nucleation. Alcohols form a surfactant layer on aqueous droplets, which can nucleate more efficiently than uncoated droplets. Ice nuclei activity has been shown to depend on alcohol chain length, molecular size and parity.[29-34] Infrared spectroscopy has been used to probe structural changes in alcohol layers during freezing,[32] and a hysteresis in ice nucleation temperature has been observed for single droplets cycled through multiple nucleation events.[33] Studies of the ice nucleation efficiency of 1-nonadecanol monolayers on water also showed a slower change in nucleation rate coefficient with temperature than observed for nucleation on a solid surface.[34] This was attributed to a reduced compatibility of the alcohol monolayer with the ice embryo as the temperature decreased.

The literature that describes surface science studies of water-surface interactions on metal and metal oxide surfaces is extensive while water interactions with carbonaceous surfaces have been less frequently studied. This is particularly true above 170 K where traditional surface science techniques cannot be applied because of the high vapor pressures above ice and liquid water. Adsorption, desorption and crystallization kinetics of thin ice films on graphite have been studied in earlier work,[27, 35-38] including helium and light scattering studies of the formation of water ice on graphite between 110 and 180 K.[27] The same techniques have also been used to study the formation of water-ammonia ice on graphite[39] and to investigate the effects of adsorbed $N_2O_5$ and $HNO_3$ on ice formation.[40] Water was observed to wet graphite at 110-140 K, while three-dimensional ice structures were formed at higher temperatures.[27] Desorption of adsorbed water molecules competes with water incorporation into the ice, making the ice formation rate strongly temperature dependent. Molecular dynamics simulations of water clusters on graphite at 90-180 K showed that at low temperatures most molecules are in direct contact with the graphite surface, while at high temperatures multi-layer cluster structures are preferred.[27] In related work, grand canonical Monte Carlo simulations were used to simulate adsorption isotherms of water molecules on different types of model soot particles.[41] The initial adsorption was favoured by strongly hydrophilic sites and optimized pore structures, and the main driving force for water adsorption was the formation of new water-water hydrogen bonds with the already adsorbed water molecules.

Here we investigate the condensation of water on graphite and alcohol-covered graphite surfaces. The overall aim is to identify and characterize governing mechanisms in ice formation by deposition freezing, which can be used to guide the further development of a molecular-level description of the process. The investigated surfaces have pronounced hydrophobic and hydrophilic properties. They are simplified compared to systems found in the atmosphere, but display some characteristics that mimic properties of soot, soot coated with organics, and secondary organic aerosol particles. We describe the results from studies of pure water, and water-methanol and water-butanol mixtures at temperatures from 175 to 213 K using elastic helium and light reflection techniques. Experiments at these relatively high temperatures were made possible by the use of a recently developed Environmental Molecular Beam (EMB) method that allows for experiments at pressures up to the $10^{-2}$ mbar range.[45]

The experimental methods are described in Section 2, followed by presentation and discussion of the main results in Section 3 and concluding remarks in Section 4.

**2. Experimental**

All experiments were performed in the recently developed EMB apparatus, which has been described in detail elsewhere.[27,42-45] The apparatus consists of six differentially pumped vacuum chambers and the main components are schematically illustrated in Figure 1. The molecular beam is generated by a pulsed gas source with part of the gas passing through a skimmer to form a directed low-density beam. A rotating chopper in a second chamber selects the central portion of each pulse producing square-wave-like frequency modulated 400 μs beam pulses. The beam source is run with pure helium at 2 bar, which produces a He beam with a mean kinetic energy of 64 meV after the chopper. After passing through a third differentially pumped chamber the beam enters the main ultra-high vacuum (UHV) chamber. The UHV chamber has a background pressure of approximately $10^{-9}$ mbar primarily due to residual background gases introduced during experiments. The beam is directed towards a surface of highly oriented pyrolytic graphite (HOPG, produced by Advanced Ceramics Corp., grade ZYB) located in the center of the UHV chamber. The 12 × 12 mm HOPG surface is cleaned between experiments by heating to 600 K, and surface conditions are routinely confirmed by elastic helium scattering after surface cooling to 200 K or lower. The UHV chamber is equipped with a rotatable differentially pumped quadrupole mass spectrometer (QMS) that can be used to measure both the incident beam and the flux from the surface.

In the EMB configuration the HOPG surface is surrounded by a separate inner environmental chamber (Figure 1) that allows for experiments with vapor pressures in the $10^{-2}$ mbar range. The finite pressure distinguishes the method from traditional molecular beam experiments, and it has been termed EMB in analogy with environmental scanning electron microscopy. The apparatus has been designed to keep pressure differences along the surface plane below 1%, while minimizing the molecular beam path length (28 mm) within the high-pressure zone.[45] The attenuation of the helium beam due to gas collisions within the inner chamber is significant at vapor pressures above $10^{-3}$ mbar, necessitating that the measured helium intensity be corrected for its attenuation based on independent measurements of beam transmission as a function of vapor pressure.[45] During studies of vapor deposition on the cold graphite substrate, water and methanol gas are introduced into the inner chamber through separate gas inlets. The incident He beam enters the innermost chamber through a circular opening with a diameter of 5 mm and collides with the surface at an incident angle of 45°. The outgoing flux passes through a second 5 mm opening in the inner chamber wall and is monitored with the QMS at a reflection angle of 45°. The QMS is also used to monitor the flux of water and methanol from the inner chamber.

In the sub-monolayer regime surface coverage of water and methanol are probed using elastic helium scattering. Surface scattering of helium atoms is highly sensitive to adlayer coverage and it has been extensively applied in surface science studies,[46,47] including studies of ice formation[27,39,40,45] and ice surface properties.[44,48] The helium scattering intensity $I_{He}$ from clean graphite surfaces is relatively intense at low temperatures but almost completely attenuated by adsorbed ice layers,[27,39,40,45] and the effective helium scattering cross section scales to a good approximation with the surface coverage when islands grow on the surface.[47] Thus the measurements can be used to systematically monitor the surface coverage as a function of time during deposition experiments. In addition to elastic helium scattering, ice layer thickness on the length scale of tens of nanometers is probed by laser reflection measurements.[27] The wavelength (670 nm) and low power (860 μW) of the diode laser ensure that heating due to absorption is negligible. At this wavelength the absorption coefficients for ice[49] and graphite[50] imply transient heating effects many orders of magnitude below the resolution of the experimental thermometry. The laser beam is directed through a quartz window at the HOPG surface with a 3° incidence angle and the reflected intensity $I_R$ is continuously recorded with a photo diode. The signal can be interpreted in terms of ice layer morphology and thickness.

## 3. Results and discussion

Here we present measurements of water and alcohol ice formation on graphite with surface temperatures $T_s$ between 175 and 213 K. Experiments with pure water are described first (Section 3.1) followed by the results for co-adsorption of methanol and water in Section 3.2 and *n*-butanol and water in Section 3.3.

## 3.1 Deposition freezing of pure water on graphite

Studies of water condensation on graphite were performed to investigate the initial nucleation and growth of pure ice on the graphite surface and to evaluate the effect of growth rate on ice properties. Figure 2 shows examples of the normalized helium scattering intensity $I_{He}$ and normalized reflected light intensity $I_R$ as a function of time during deposition freezing experiments from $T_s$ = 185 K to 213 K. At high $T_s$ the relative background contribution to $I_{He}$ becomes larger because the probability of elastic helium scattering from clean graphite decreases.[46,47] In each experiment the water source was turned on (indicated by shading, Figure 2) and the vapor pressure adjusted to attain a critical supersaturation for ice nucleation. Subsequently, the $H_2O$ vapor pressure decreased to a lower steady-state level as gas phase $H_2O$ was deposited as ice. The steady-state water vapor pressures increased rapidly with increasing temperature and reached $1.9 \cdot 10^{-2}$ mbar at 213 K. At each temperature the $I_{He}$ decreased slowly during a short initial ice nucleation phase followed by a steep decrease until a substantial fraction of the surface was covered with ice. From this point $I_{He}$ again decayed more slowly until the graphite surface was completely covered. At 185 K, 195 K, and 205 K (Figure 2(a)-2(c)) water ice grew to cover the graphite surface completely, thus $I_{He}$ was observed to reach a constant minimum value. At 213 K (Figure 2(d)) the water inlet was turned off before the surface was completely covered. In each case after the water inlet was turned off the ice desorbed and the graphite returned to the original bare surface conditions. The time required for full desorption was primarily determined by the evaporation rate. In general, evaporation results in non-negligible partial water vapor pressures in the inner chamber, allowing some material to recondense and extending the total evaporation time.

In Figure 2 the initial decrease in $I_{He}$ is followed by a decrease in light scattering intensity. That indicates ice structures with substantial thickness are produced even before the graphite surface is completely covered by ice. As the ice layer forms, $I_R$ is attenuated with time much like $I_{He}$. The attenuation of the signal is overlaid by a sinusoidal-like time variation that is produced by the constructive and destructive interference of light reflected from the growing ice layer and the underlying graphite. The light scattering pattern observed later as the ice evaporates mirrors the growth pattern, suggesting that ice formation and removal is largely a reversible process with limited internal relaxation of the structure during the lifetime of the layer. Based on the combined results from helium and light scattering experiments we conclude that > 10 nm thick ice existed simultaneously with bare graphite patches, indicating that water did not efficiently wet the surface.

Additional experiments demonstrated that the ice growth rate and structure depended significantly on the initial H$_2$O vapor pressure. This is illustrated in Figure 3 where ice formation at 195 K was monitored for three different initial vapor pressures. The helium scattering data illustrates that the time required to reach complete surface coverage decreases considerably as the initial pressure increases. The corresponding light reflection data indicates that the morphological properties of the ice also change. Slow ice growth results in ice structures with substantial thicknesses before the graphite surface is fully covered with ice, while higher initial vapor pressures produce relatively flat and transparent ice as revealed by the light scattering pattern.

The evolution of the ice morphology can be quantitatively characterized using a surface roughness model to physically interpret the light reflection data. The reflected light intensities $I_R$ are observed to partially attenuate and oscillate in time. The periodic behavior is the expected result of wave interference due to multiple reflections within the ice layer.[51] The beam attenuation requires a more complicated explanation, which we propose is well explained by ice surface morphology.

Figure 4(a) illustrates an idealized experimental geometry of a smooth ice layer on graphite. The waves of experimental interest are the initial reflection from the ice surface and subsequent reflections from the ice/graphite interface. For isotropic materials this Fabry-Pérot type of multiple wave interference is quite straightforward[52] but becomes quite complicated with the introduction of anisotropic materials like ice and graphite.[51] Fortunately, ice is only weakly anisotropic and reflection from the graphite only occurs from a single faceted surface. Thus, for practical purposes we are able to use the isotropic approximation, for which the ice layer thickness $d$ can be related to the intensity maxima (minima) resulting from constructive (destructive) interference,

$$m\lambda = 2n_i d \cos\theta_{ice}, \tag{1}$$

$$\left(m - \frac{1}{2}\right)\lambda = 2n_i d \cos\theta_{ice}, \tag{2}$$

where $m\lambda$ is an integer number ($m = 1,2,3…$) of wavelengths with $\lambda = 670$ nm, $n_i$ is the index of refraction of ice, and $\theta_{ice}$ is the angle of propagation within the ice layer computed from Snell's Law. Here because we are unable to determine the ice orientation we take $n_i = 1.3097$ to be the average of the ordinary and extraordinary indices of refraction.

Equations 1 and 2 allow us to transform $I_R$ intensity measurements into time evolving ice layer thicknesses. However a more detailed analysis is needed to explain the observed signal attenuation. Previous studies have shown that as ice forms on graphite it can form islands and plateaus resulting in a corrugated surface.[35] This suggests that our idealized ice layer can be better approximated with a rough surface (Figure 4(b)). Beckmann et al.[53] showed that for normally distributed roughness, reflected intensity in the specular direction is scaled with a parameter $S_r$:

$$S_r = exp\left[-\left(\frac{4\pi\sigma cos\theta_i}{\lambda}\right)^2\right], \quad (3)$$

where $\sigma$ is the standard deviation of the roughness and $\theta_i$ is the incident angle. Here the roughness is considered as a departure from a mean height, thus $\sigma$ is a measure of the amplitude of the surface roughness.

For an exact theoretical calculation the roughness scattering coefficient $S_r$ would have to be accounted for in each successive reflection. One would also expect that for such surfaces another amplitude adjustment is necessary for the transmitted beam. However, in this case we capture the fundamental behavior simply by scaling the entire reflected intensity ratio

$$I'_R = S_R I_R. \quad (4)$$

Although this is not an exact theoretical solution it allows for a straightforward comparison with the experiment.

The combined theoretical construct of multiple wave interference and surface roughness allows us to simultaneously constrain ice layer thickness, growth rate, and surface roughness. Using the normalized and digitized constructive interference maxima as the scaled amplitudes, we can solve Eq. 4 for $\sigma$ yielding estimates of the time evolution of the surface roughness.

This technique is used in Figure 5 to constrain the time evolution of ice thickness $d$ and roughness $\sigma$ obtained from $I_R$ at $T_s$ = 185 K and 195 K, corresponding with the experimental data shown in Figure 2. Roughness increased during initial growth until the surface became completely covered. After a complete layer was formed roughness was observed to remain constant even as the ice layer thickened. Calculated surface roughnesses between 25 and 60 nm, are much less than the maximum ice thicknesses, which are more than one micron. Measured roughnesses for all temperatures and different nucleation vapor pressures lie within this range, illustrating that the $\sigma$ is a complex function of the pressure and temperature. Thus within this range of temperature, obtaining smoother surfaces appears to require vapor pressures that strongly favor deposition. In this way adsorbing molecules overwhelm the outgoing flux and build more complete layers. However, at high temperatures such high vapor pressures are difficult to sustain for entire experiments and rough surfaces generally result. It is important to note that due to the assumptions of the simplified model it necessarily predicts upper limits of $\sigma$. If other factors like polycrystallinity that potentially increase light attenuation could be treated completely the effect would be to decrease the measured roughness. In that regard the results are also limited by oscillations in the power of the laser source, which leads to an additional uncertainty in $\sigma$ of 7 nm. Conversely, the ± 10 nm uncertainty in the calculated ice thicknesses is relatively much smaller and only due to the refractive index assumption and the uncertainty in identifying interference maxima/minima. The merit of the simple model is that it enables unique simultaneous measurements of bulk morphologic properties and molecular kinetics. Furthermore, the limited available data of homologous systems suggests the growth of crystal domains representative of the roughnesses we have measured.[35]

For pure water ice these results are also consistent with an earlier study in which three-dimensional ice structures were formed between 140 and 180 K, and planar ice layers were formed at lower temperatures.[27] The differences highlight the importance of intermolecular interactions in the adsorbed layer relative to weaker water-surface interactions, in agreement with earlier studies of water interactions with graphite[27,36-40,54,55] and other substrates.[56,57] In the sub-monolayer regime initial ice growth is characterized by a competition between the rapid desorption of individual $H_2O$ molecules from bare graphite and their diffusion and incorporation into existing ice structures. For small initial surface aggregates the time scale of ice growth is similar to the time scale for internal structural reconstructions. Thus there exists a sensitive balance between surface morphology and ice layer thickness. These observations of pure water indicate that on the time scale of minutes this balance is established by the initial conditions of ice nucleation and growth. At these temperatures, especially below 200 K, relaxation time scales maintain corrugated ice surfaces even as the ice depth precludes influence of the graphite surface. This conclusion has potential relevance for the properties of ice clouds in the upper troposphere, where growth of frozen particles may result in sedimentation and relatively short particle lifetimes.[58]

## 3.2 Co-adsorption of methanol and water

Introducing methanol had distinct effects on the formation and growth of ice layers. Figure 6 illustrates a typical freezing experiment at 185 K when water was being inlet from 35 to 410 s, and methanol was introduced simultaneously from 70 to 250 s. During the initial stage of the experiment, before methanol was introduced the water vapor pressure was $4.2 \cdot 10^{-4}$ mbar. Although this exceeds the equilibrium pressure of ice ($2.4 \cdot 10^{-4}$ mbar) at 185 K[59-62] the supersaturation of 75% was insufficient to generate rapid ice nucleation on the bare graphite surface before the introduction of methanol. Subsequently, the $CH_3OH$ pressure was increased in steps, reaching a maximum value of $4 \cdot 10^{-3}$ mbar after 100 s. With methanol present the graphite surface coverage increased rapidly with increasing methanol pressure indicating that lateral interactions play an important role for the adsorption enthalpy.[45] This agrees with earlier work which showed that under these conditions methanol rapidly condenses and forms a thin layer on graphite.[45,63] The attenuation of $I_{He}$ was quickly followed by a drop in water vapor pressure, due to uptake, and the appearance of interference oscillations in $I_R$, which are both indicative of a thick and growing $H_2O$-containing ice layer. Furthermore the lack of attenuation of the $I_R$ interference maxima signifies that the ice layer was smooth and transparent. This is in sharp contrast to the pure water ice growth (Figure 2) where rough surfaces were formed. Thus the presence of methanol has a strong impact on both the ice nucleation and morphologic structure. As the ice grew the methanol pressure gradually decreased suggesting that some methanol was being incorporated into the ice. After the methanol inlet was turned off at 247 s the $I_R$ intensity shows that smooth ice continues to grow until 410 s when the water inlet was also closed. Immediately the ice began to desorb reversing the interference oscillations and returning to a bare graphite surface by 700 s. During this time the surface evaporation maintained a relatively high water vapor pressure but did not produce a major methanol release. Thus within the ice the methanol concentration must have been low.

To further investigate the effect of methanol on the deposition freezing process, experiments were carried out with different initial methanol coverages prior to water exposure. The results at 185 K are summarized in Figure 7. First, as an experimental control pure ice is nucleated and grown (c.f., Figure 2-3) from a high initial vapor pressure, in order to minimize the surface roughness (Figure 7(a)). In subsequent experiments methanol layers of 100%, 20%, and <10% coverage are added before introducing H$_2$O (Figure 7(b)-(d)). In each case the unattenuated $I_R$ shows that the addition of methanol results in a smoother ice structure than is observed for the pure case. As Figure 7(d) illustrates, even low methanol coverages are sufficient to impact the structure of the frozen layer. The effect is quantified in Figure 8 where the surface roughness calculated from the light reflection theory is plotted as a function of alcohol coverage. Reversing the order of deposition by first nucleating ice on the graphite surface and then adding methanol reduces the observed methanol effect. No effect is observed when methanol is added to a completely covered ice surface (Figure 9(a)). Adding CH$_3$OH to partially ice-covered surfaces (Figures 9(b) and (c)) results in rapid layer growth to complete surface coverage (fully attenuated $I_{He}$) and a change in thickness growth rate indicated by the $I_R$. The change of growth rate and reduced rate of $I_R$ attenuation also suggest that the growing ice incorporates some methanol. However, these effects are small relative to those cases when methanol was first condensed on the surface. In each case the final result is a rough ice morphology, similar to what is formed in the pure case, whose growth is negligibly affected by the presence of methanol.

The reason that even an incomplete adlayer of methanol has such a profound effect on the nucleation of ice and structure is rooted in the diffusional time scales and intermolecular interactions of the system. In general the methanol-graphite interaction is stronger than both the water-graphite interaction and the self-attraction of methanol.[63-65] Thus the methanol preferentially forms a monolayer on graphite, while the formation of multilayers is limited by comparatively weak methanol-methanol interactions.[63,64] Thomson et al.[63] recently studied water adsorption to these thin methanol layers and showed that although most impinging water molecules quickly desorb from such layers some are incorporated into the methanol layer. This incorporation of water molecules provides a stable H$_2$O repository during the initial ice nucleation phase and thereby facilitates ice formation. Although the system is always highly dynamic with constant high rates of adsorption and desorption, a stable methanol monolayer increases the H$_2$O surface residence time enabling ice structures to form and spread across the entire surface. Conversely, nucleation resulting from deposition of H$_2$O on clean graphite is a statistical thermodynamic process that results in individual ice domains[27,35]. Because the self-attraction of water is stronger than its attraction to the graphite these individual ice islands grow and only cover the surface when their roughness is comparable to their initial separation. In these cases the ice surface roughness is preserved due to the long relaxation time scale associated with H$_2$O self-diffusion. The length scale of diffusion $L(H_2O) \propto \sqrt{Dt}$, where $D$ is the diffusion coefficient and for these experiments $t$ = 100-1000 s can be estimated using previous results for $D = 10^{-16}$-$10^{-18}$ m$^2$s$^{-1}$.[66] The result $L$ ~10 nm is smaller than the observed roughness, thus it is not surprising that little surface smoothing is observed. Here it is also relevant to note that at these temperatures diffusion of methanol in ice is up to an order of magnitude faster than the self-diffusion of water.[66]

Methanol also acts as a surfactant and concentrates on the ice surfaces.[67-70] The veneer of methanol decreases the surface tension and enhances the adsorbate-graphite interactions. Simulations of ethanol surfactants on water clusters have modeled this effect as a reduction in the cluster-graphite contact angle.[71] Although methanol primarily resides on the ice surface,[70] our results suggest that some methanol is also incorporated into the growing ice structure. This is noted in Figure 6 where the methanol pressure decreases above the growing ice surface. After the methanol inlet is turned off the methanol pressure quickly drops before assuming a slow decay during the remaining monitoring period. Figure 10 shows a detailed view of the methanol pressure decay during the transition from ice growth to evaporation. A step increase interrupts the steady decay when the water inlet is turned off and the ice layer begins to decrease in thickness. Because the surface-vapor exchange of methanol is temporally continuous this observation must result from the enhancement of bulk diffusion of methanol and subsequent desorption during ice sublimation. Importantly, the background methanol pressure was sufficient to maintain substantial methanol coverage on the ice throughout the experiment,[70] and thus the desorbing flux was primarily due to excess methanol reaching the outer surface from within the bulk. Using Arrhenius parameters reported by Marchand et al.,[72] and Livingston et al.,[66] respectively, the methanol diffusion length $L_M$ in ice for t = 100 s at 195 K can be estimated, $30 < L_M < 750$ nm. This broad range and the uncertainty in diffusion coefficient have previously been ascribed to contrasting ice structures in different studies and to a large effect of methanol concentration on the diffusion rate.[70] The upper limit of these values is consistent with the rapid methanol release observed in these experiments and thus the conclusion that the methanol fraction incorporated into the ice is finite but low.

A small $CH_3OH$ peak was also observed during the final stages of the ice evaporation indicating that near the graphite surface a high methanol concentration was maintained throughout the experiments. This is illustrated in Figure 11 where the surface coverage $\Theta$ is plotted together with $I_R$, the water intensity $I_W$, and the methanol intensity $I_M$ as the graphite re-emerges in two experiments. The surface coverage, $\Theta = (1 - I_{He}) / (1 - I_{bk})$, where $I_{bk}$ is the steady state He intensity reached for a completely ice-covered graphite surface and includes contributions from the background level, is introduced to simplify comparison with the other quantities. In the pure water case shown in Figure 11(a), $I_W$ and $\Theta$ decrease simultaneously as the bare graphite surface re-emerges. This is expected because the desorbing flux is related to the available ice surface area. In co-adsorption experiments (Figure 11(b)), the final decrease in $I_W$ is associated with an increase in $I_M$ indicating a change in chemical composition of the ice closest to the graphite. Furthermore, bare graphite is not exposed until $I_W$ has substantially decreased. Similar results as in Figure 11(b) were obtained for $H_2O$-$CH_3OH$ layers produced under different conditions. The fact that the ice layers remained thick as the methanol emission increased is indicative of a methanol concentration gradient as a function of distance from the graphite surface. However, the shrinking adlayer is a relatively dynamic system and as previously mentioned rapid methanol diffusion likely plays a role in the methanol emission. Ultimately the experiments with methanol illustrate two areas of increased $CH_3OH$ concentration, the first near the graphite surface due to the strong methanol-graphite interaction, and the second on the $H_2O$ surface. Between these surfaces methanol does appear intermixed within the bulk ice in low concentrations.

### 3.3 Co-adsorption of *n*-butanol and water

Additional deposition freezing experiments were carried out with *n*-butanol replacing methanol. In Figure 12 results at 185 K are presented. Butanol was inlet from 80 to 420 s, with water being introduced between 185 and 490 s. Initially a layer of butanol covers the surface, attenuating $I_{He}$, but the lack of simultaneous $I_R$ attenuation indicates only thin layer formation. This is consistent with earlier studies, which showed that *n*-butanol forms a stable monolayer on graphite.[73] When the water inlet is turned on a thick ice layer begins to grow.

Both the water and butanol pressures decrease during ice growth as both gases are simultaneously incorporated into the ice structure. At 420 s the butanol inlet is closed but steady ice growth continues, indicating that butanol is only a minor component in the ice. When the water inlet is also closed (490 s) the ice layer begins to thin, maintaining a finite chamber water pressure until the ice is fully evaporated. When the ice evaporates the butanol pressure increases as it is released from the ice. As the final layers of ice evaporate a transient increase in butanol pressure suggests that the layers closest to the graphite are enriched in butanol, similar to the methanol case. Finally, a thin butanol layer remains on the graphite surface due to its relative stability.[73] Calculated surface roughnesses from $I_R$ span an identical range to measurements of the pure water case, quite distinct from the smoother methanol-water mixture results (Figure 8). The inability of butanol to initiate structural changes in the ice surface makes it apparent that in this context its properties contrast sharply with those of methanol.

Adsorption of butanol onto previously formed water ice layers at 185 K did not appear to change the properties of the growing ice on the time scale of a few hundred seconds. However, light reflection data revealed that ice at 185 K evaporated 39% slower with a butanol-layer present compared to the bare water ice case. Thus the butanol adlayer influences water desorption, in sharp contrast to adsorbed methanol that has no observable effect. Previously, long-chain alcohols have been found to strongly reduce evaporation from liquid water[74,75] and recent MD simulations suggest that butanol, hexanol and octanol should have a similar but reduced effect.[76] The present results for the butanol-water system are consistent with these earlier studies.

## 4. Concluding remarks

We have investigated the formation of ice by deposition freezing on bare and alcohol-covered graphite surfaces at temperatures from 175 to 213 K. The studies were made possible by the recent development of the EMB technique that allowed probing of surface conditions with elastic helium scattering at vapor pressures up to $10^{-2}$ mbar above the surface, simultaneously with reflected light intensity measurements. The technique enables highly dynamic surfaces to be categorized in terms of surface coverage, morphology, and adsorption/desorption rates.

Bare graphite is an inefficient substrate for ice nucleation due to the weak water-graphite interactions. Water has a relatively low wettability on graphite and deposition results in growth of rough ice surfaces with estimated surface roughnesses from 25 to 50 nm that depend on growth rate and temperature. Adding methanol to the system has a profound effect on the ice nucleation process. Adsorbed methanol provides a hydrophilic substrate for ice nucleation and stabilizes water structures during nucleation by the formation of hydrogen bonds between adsorbed water and methanol molecules. Pre-adsorption of a methanol monolayer results in the growth of a smooth crystalline ice phase and pronounced structural effects are observed even at initial surface coverages of less than 10%. Contrastingly, once the substrate is completely ice covered the presence of methanol has a negligible effect on the continued ice growth. Unlike with methanol, ice formation on *n*-butanol-covered graphite results in rough ice surfaces suggesting that water ice does not efficiently wet the butanol layer.

Both methanol and butanol primarily reside on the ice surface and at the ice-graphite interface, with minor fractions incorporated into the growing ice and released by diffusion on the time scale of the experiments. Deduced rates of diffusion are within the upper limits of previously published values and suggest that it is likely that the ice is highly polycrystalline, and thus grain boundaries and other regions of enhanced mobility facilitate the diffusive process. The alcohol surfactants may also have a significant effect on the ice growth rates. While methanol does not appear to have a measurable effect on water uptake, the adsorption of butanol at 185 K results in a reduction in the ice evaporation rate. Quantifying the efficiency of butanol as a transport barrier for adsorption is experimentally more difficult but is expected to mirror its ability to hinder desorption.

Interestingly, for all three investigated systems the ice structure is mainly determined by the conditions in the sub-monolayer regime where growth and internal relaxation take place with similar rates and the sub-monolayer structure is largely determined by the strength of the substrate/adsorbate interactions. The hydrophilic methanol-covered surface produced clear and laminar crystalline ice, while the hydrophobic bare and butanol-covered graphite result in rough ice surface structures. The initial ice structures are conserved during subsequent ice growth up to micrometer thicknesses and sublimation is largely a reversible process with limited structural relaxation on the time scale of minutes used in the present studies. This observation may be relevant in atmospheric conditions where ice particles grow with typical time scales from minutes to hours. Thus the surface structure of cirrus ice particles may maintain a signature of the initial ice growth conditions over extended time periods.

Adsorbed short-chain organic molecules can have significant effects on the ice nucleation process and on the structure and growth of ice formed by deposition freezing. The investigated systems display characteristics ranging from hydrophobic in the case of bare and butanol-covered graphite to hydrophilic in case of methanol-covered graphite, and the hydrophilicity is observed to have a significant effect on the ice nucleation process by stabilizing and changing the structure of ice embryos on the surface. Although, the alcohol pressures used here are higher than encountered in the atmosphere, and methanol is not believed to play an important role in the atmosphere because of its relatively low stability on carbon surfaces, similar compounds with low vapor pressures that provide suitable sites for strong water interactions may have similar effects. Previous experiments with particles formed by different combustion processes have shown that soot particles may act as relatively efficient IN due to the presence of hydrophilic surface groups that can form strong bonds with water molecules.[20-25,28] Although such groups are often inorganic compounds or chemically bound to the carbon structure, the present study illustrates that adsorbed organic compounds with a high O/C ratio may also modify the IN efficiency and provide surface sites that maximize the formation of hydrogen bonds and enhance growth of ice embryos on the surface. Studies of common stable products of the oxidation of atmospheric organics are planned to further explore the effect of organic compounds on ice nucleation in the deposition mode.

## Acknowledgement

Funding for this research was provided by the Swedish Research Council and the University of Gothenburg. Mr. Benny Lönn is gratefully acknowledged for technical support.

Kong *et al.* Figure 1

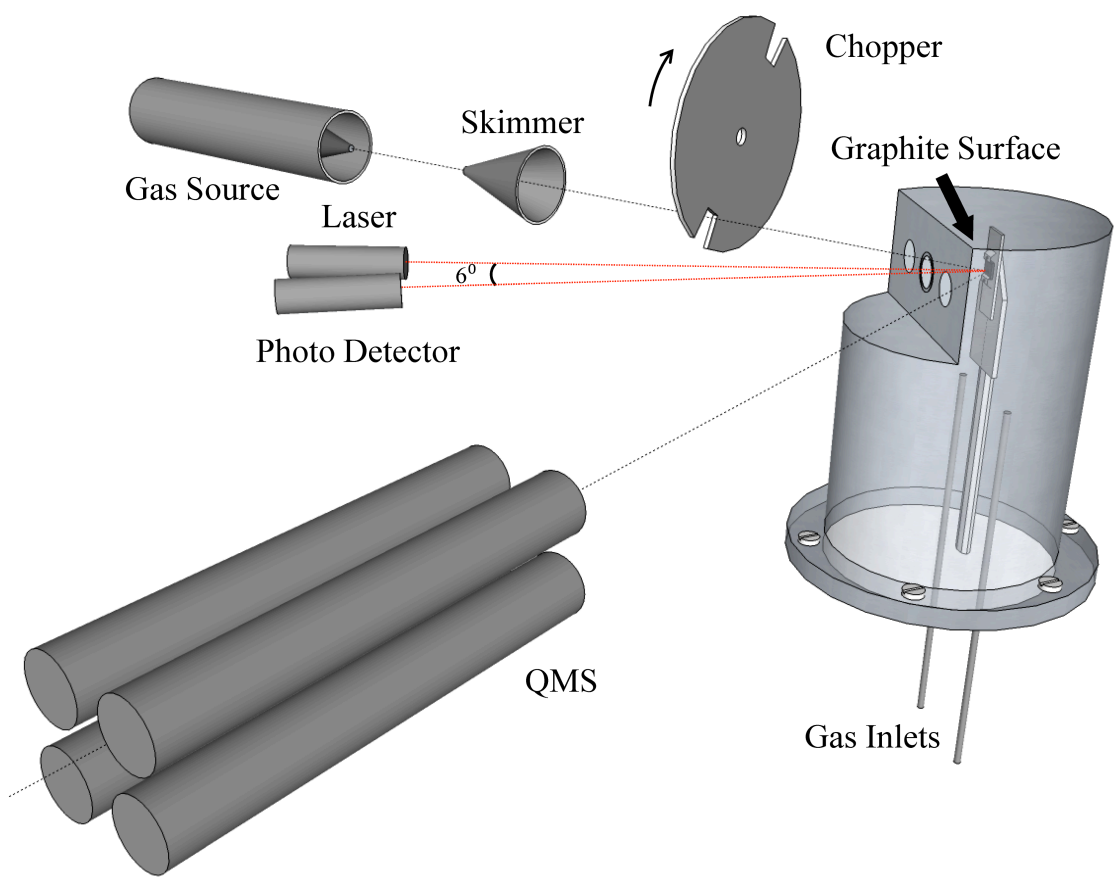

**Figure 1.** Schematic view of the environmental molecular beam apparatus including the main components and the inner chamber surrounding the graphite surface.

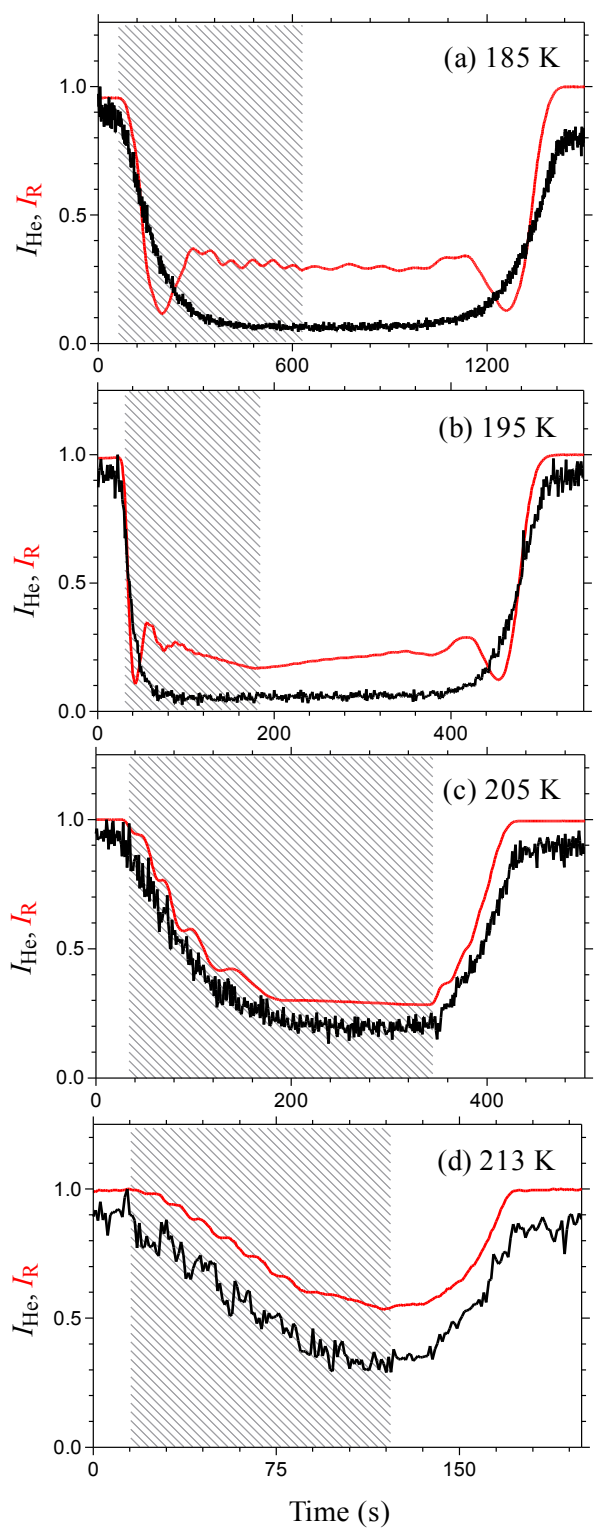

**Figure 2.** Water ice formation on graphite from 185 to 213 K: normalized scattered helium intensity $I_{He}$ and reflected light intensity $I_R$ as a function of time. The shadowed areas indicate when the water inlet is on. Surface temperatures are indicated in the panels. The water pressures in the inner chamber when a complete ice layer had formed were $4.8 \cdot 10^{-4}$ mbar (185 K), $2.3 \cdot 10^{-3}$ mbar (195 K), $7.3 \cdot 10^{-3}$ mbar (205 K), and $1.9 \cdot 10^{-2}$ mbar (213 K).

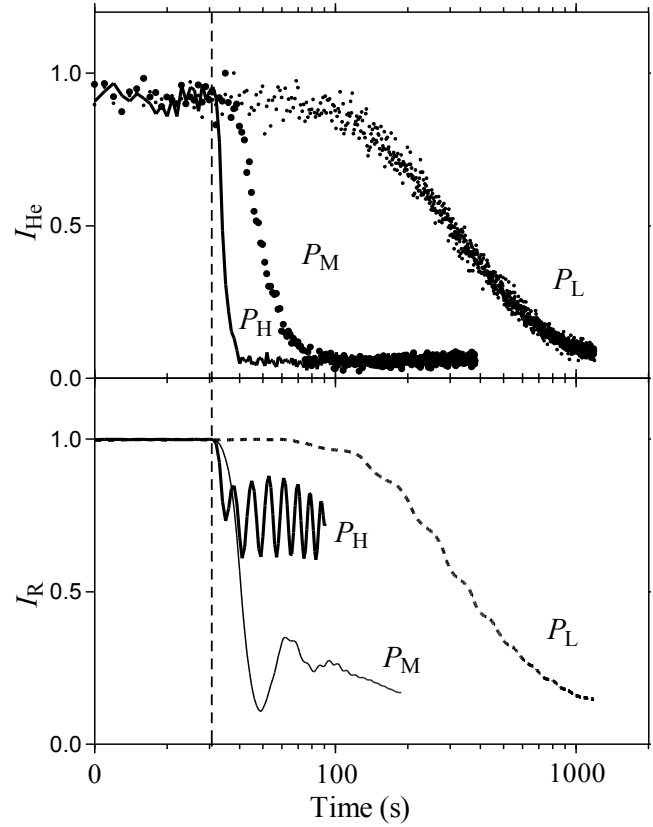

**Figure 3.** $I_{He}$ and $I_R$ as a function of time during ice formation on graphite exposed to different water vapor pressures at 195 K. Vapor pressures; $P_L = 1.4 \cdot 10^{-3}$ mbar, $P_M = 1.7 \cdot 10^{-3}$ mbar and $P_H = 2.0 \cdot 10^{-3}$ mbar, were recorded after reaching a steady state.

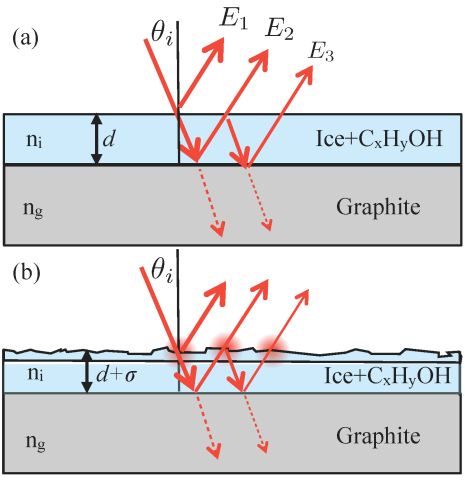

**Figure 4.** Schematics of (a) idealized multiple specular reflections from ice covered graphite, and (b) scattered reflections from a rough ice surface.

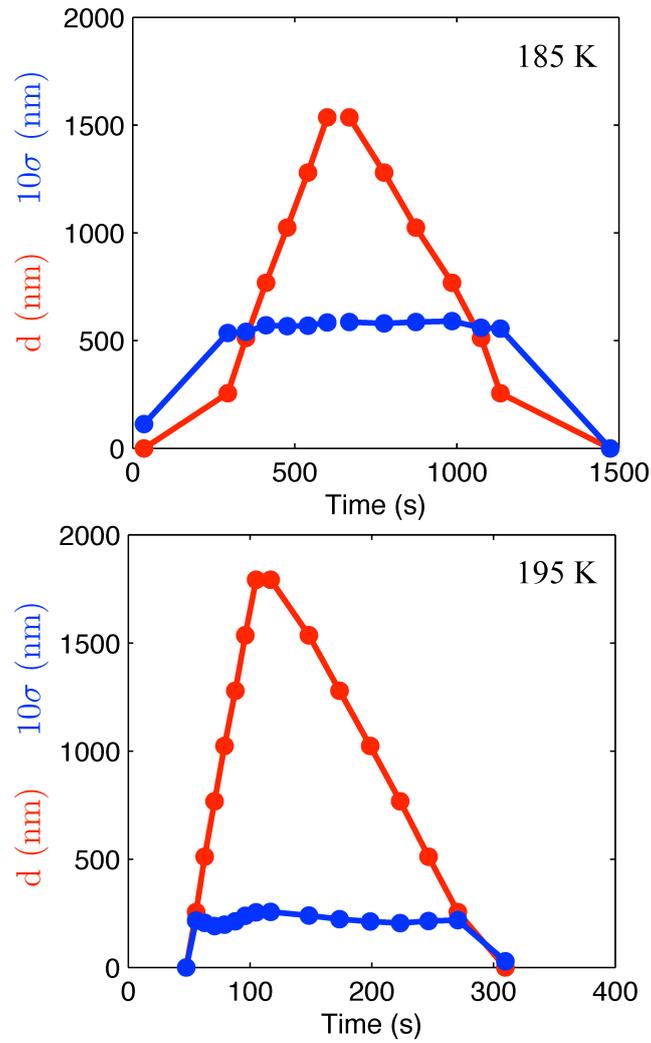

**Figure 5.** Time evolution of ice surface thickness *d* and roughness $\sigma$ at (a) 185 K and (b) 195 K, calculated using the model described in Section 3.1. For a discussion of the uncertainties in the calculation refer to Section 3.1.

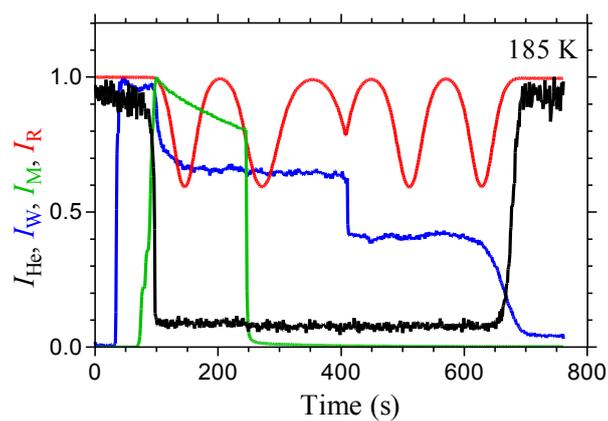

**Figure 6.** Ice formation on methanol-covered graphite at 185 K: methanol ($I_M$), water ($I_W$), scattered helium ($I_{He}$), and reflected light ($I_R$) intensities as a function of time. The water vapor partial pressure before methanol was introduced was $4.2 \cdot 10^{-4}$ mbar, and a maximum methanol vapor pressure of $5.5 \cdot 10^{-3}$ mbar was reached before ice formed on the surface.

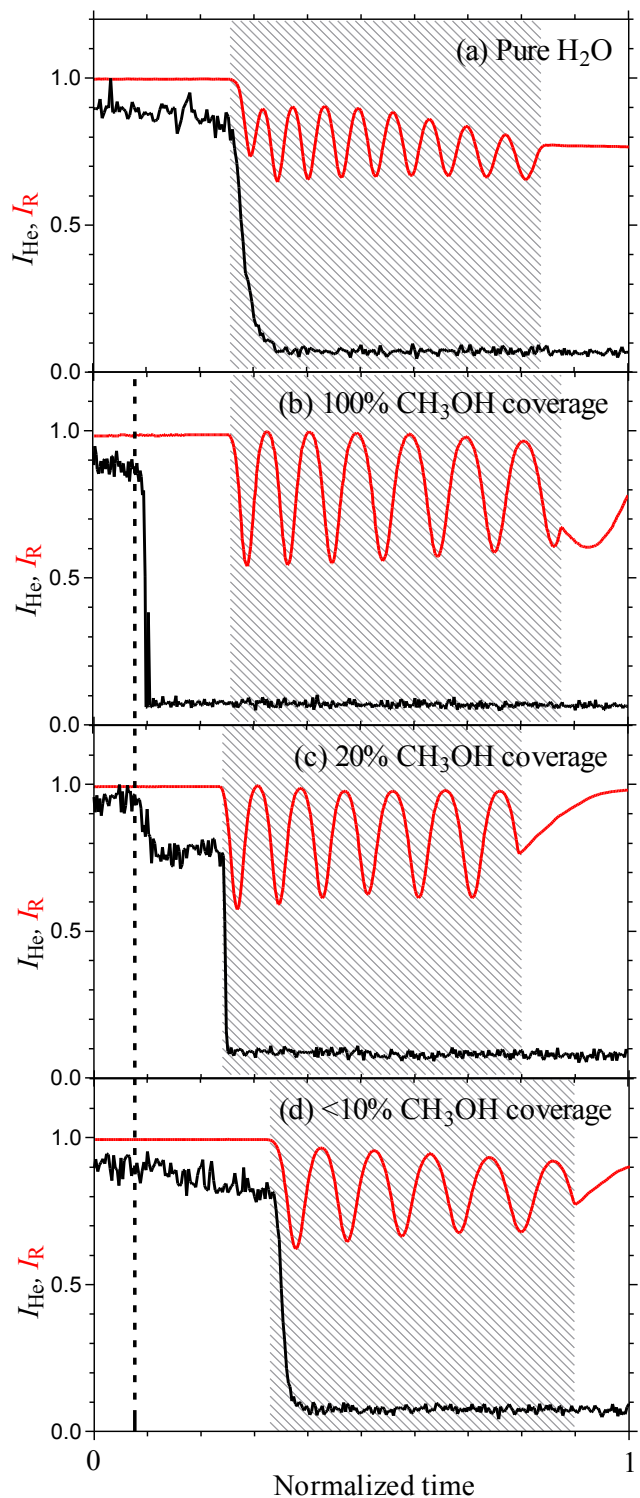

Kong *et al.* Figure 7

**Figure 7.** Ice formation on graphite with different initial methanol coverages at 185 K: $I_{He}$ and $I_R$ as a function of normalized time. The shadowing indicates when the water inlet was on and the dashed line indicates when the introduction of methanol began. The total time depicted ranged from 250 to 450 s.

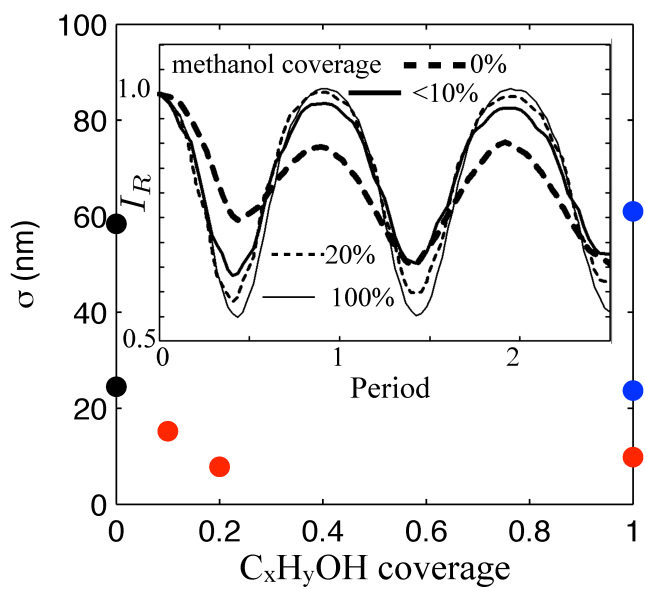

**Figure 8.** Calculated ice surface roughnesses (refer to Section 3.1 for a discussion of the calculation uncertainty) on graphite with different initial alcohol coverages: no alcohol coverage (black points), methanol coverage (red points), and butanol coverage (blue points). Examples of the first three $I_R$ intensity maxima (from Figure 6) with each oscillation period normalized to compensate for small differences in growth rate are inset to demonstrate their relative attenuations.

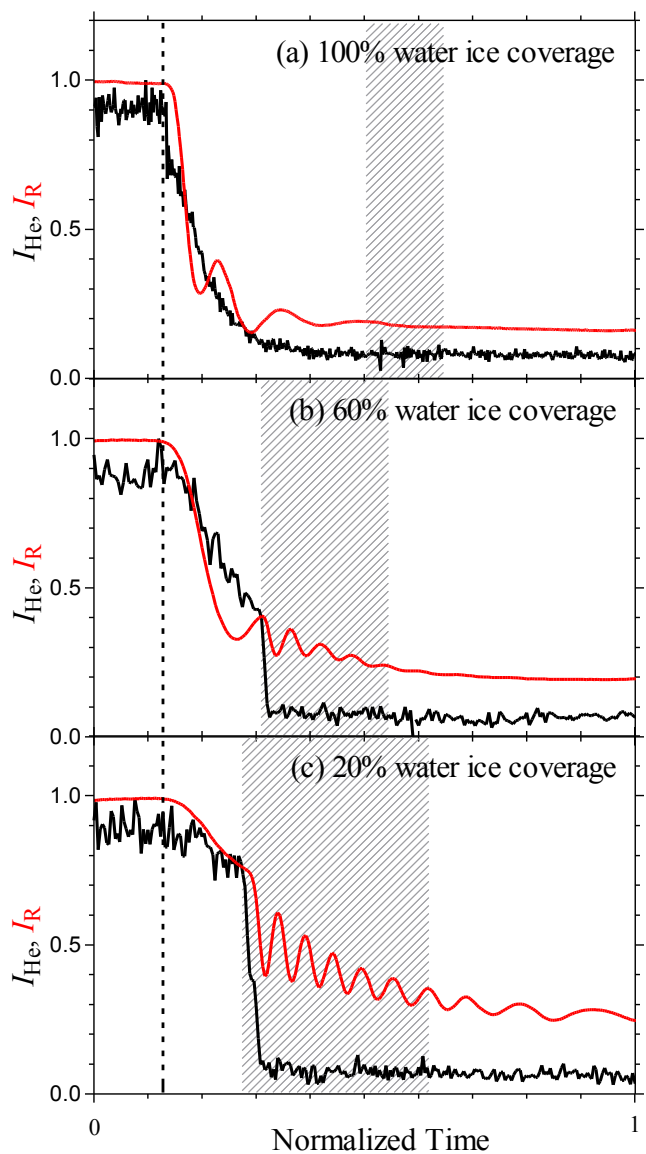

**Figure 9.** Effect of methanol inlet during formation and growth of water ice on graphite at 195 K: $I_{He}$ and $I_R$ as a function of normalized time. The shading indicates when the CH$_3$OH inlet was on with the opening of the H$_2$O inlet indicated by a dashed line.

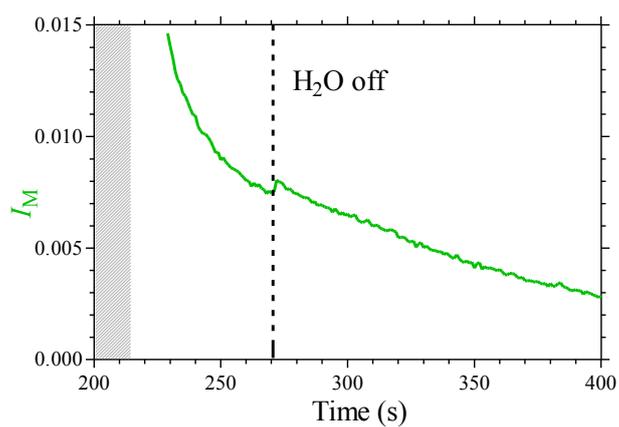

**Figure 10.** Normalized methanol intensity $I_M$ as a function of time during sublimation of an ice layer grown on methanol-covered graphite at 195 K. The time period when the water and methanol gas inlets were both on (gray area) and when the water inlet was turned off (dashed line) are indicated.

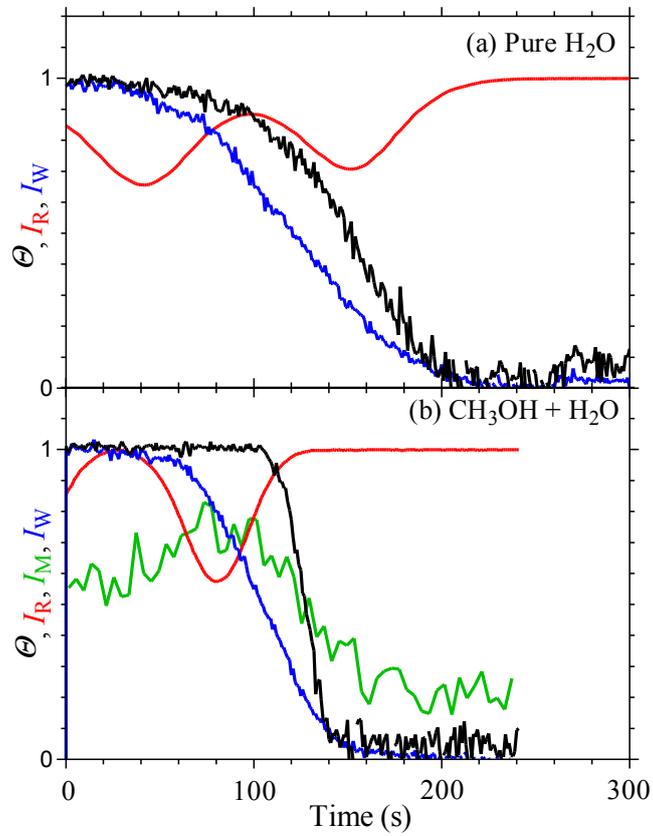

**Figure 11.** Final stage of evaporation of ice formed from (a) pure water and (b) co-adsorbed water and methanol: surface coverage $\theta$, $I_W$, $I_M$ and $I_R$ as a function of time. The results correspond to the cases shown in Figure 6(a) and Figure 6(b) respectively.

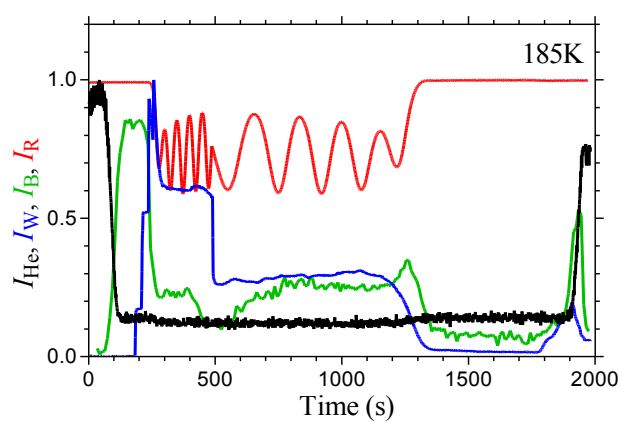

**Figure 12.** Ice formation on *n*-butanol-covered graphite at 185 K: $I_W$, $I_{He}$, $I_R$ and butanol ($I_B$), intensities as a function of time.